\documentclass[preprint,floats,aps,showpacs]{revtex4}
\usepackage{epsfig}

\def\be{\begin{equation}}
\def\ee{\end{equation}}
\def\bdm{\begin{displaymath}}
\def\edm{\end{displaymath}}
\def\bea{\begin{eqnarray}}
\def\eea{\end{eqnarray}}
\def\nn{\nonumber\\}
\def\up{\uparrow}

\def\eps{\epsilon}

\def\s{\sigma}

\newcommand{\p}{\partial}

\newcommand{\rc}{\mbox{c}}
\newcommand{\rs}{\mbox{s}}
\newcommand{\down}{\downarrow}
\newcommand{\la}{\langle}
\newcommand{\ra}{\rangle}
\newcommand{\rd}{\mbox{d}}
\newcommand{\ri}{\mbox{i}}
\newcommand{\re}{\mbox{e}}

\newcommand{\vsigma}{\mbox{\boldmath $\sigma$}}
\newcommand{\vSigma}{\mbox{\boldmath $\Sigma$}}
\begin{document}
\title{Coulomb Blockade Regime of a Single-Wall Nanotube}
\author{ A. A. Nersesyan$^{1,2}$ and A. M. Tsvelik$^3$}
\affiliation{$^1$The Abdus Salam International Centre for 
Theoretical Physics, Strada Costiera 11, 34100 Trieste, Italy\\
$^2$ The Andronikashvili Institute of Physics, Tamarashvili 6, 380077, Tbilisi, Georgia\\
$^3$ Department of  Physics, Brookhaven 
National Laboratory, Upton, NY 11973-5000, USA}
\date{\today}

\begin{abstract}
We study a  model of carbon nanotube with  a half filled conduction band. At this filling the system is the Mott insulator. The Coulomb interaction
is assumed to be unscreened. It is shown that this allows to develop the adiabatic 
approximation which  leads to considerable simplifications in calculations of the 
excitation spectrum. We give a detailed analysis of the spectrum and  the phase 
diagram at half filling, and discuss effects of  small doping. In the latter case  
several phases develop strong superconducting fluctuations corresponding to various 
types of pairing. 

\end{abstract}
\maketitle

\section{Introduction}

Carbon nanotubes  have attracted an enormous amount of attention and generated an 
immense level of theoretical 
and experimental activity. It has been rightly pointed out that single-wall 
carbon nanotubes (SWCN) represent ideal 
one-dimensional systems and one may expect to observe here exotic phenomena characteristic 
for strong correlations in one dimension. Most of the attention has been concentrated 
on a possibility of Tomonaga-Luttinger liquid. Though such liquid is certainly a very 
interesting object, it is just one of many wonders strong correlations can produce. 

In this paper we concentrate on the physics of an armchair SWCN 
at half filling. 
The features which dominate this physics are the unscreened Coulomb interaction and  
Umklapp processes.  The importance of the unscreened Coulomb  interaction 
for nanotubes away from 
half filling was already noticed in the previous studies \cite{eg},\cite{kane}. 
At half filling some additional factors come into play. It was realized in \cite{odintsov}
that the  long-range Coulomb interaction makes the operators responsible for the 
Umklapp scattering terms relevant. Consequently, the gaps for collective excitations generated 
by such operators are not exponentially small, as it would be away from half filling, but have 
power-law dependence on the Umklapp scattering matrix elements. Such  enhancement of the 
gaps increases  chances for their experimental observation. Unfortunately, in their further 
analysis the authors of \cite{odintsov} resorted to Renormalization Group (RG) equations,  
which for systems with many fields do not provide much insight into the properties at  
the strong-coupling fixed point. RG equations also do not take into account  
the drastic difference 
between  velocities of the plasma modes and all other collective excitations. Instead of 
adding difficulties, however, the  difference in velocities leads to considerable simplifications in 
actual calculations (see \cite{levitov}). In this paper we shall exploit this feature to our 
advantage and develop an approach based on the adiabatic approximation similar to the one 
used for the problem of electron-phonon interaction. As a result we will be able to provide 
a rather detailed information about the spectrum and the phase diagram of the system. 
As will be shown below, the interplay of the long range Coulomb interaction and  Umklapp 
processes  at half filling gives rise to a phase diagram which includes several interesting 
 strongly correlated states. The system has a hidden Z$_2\times$Z$_2\times$Z$_2$ symmetry 
and these phases are conveniently classified as different  symmetry breaking patterns of 
this group.  

\section{The problem of single-channel wire at half filling}

 To warm up, let us first recall the problem of 
the unscreened Coulomb interaction for
a single chain (or a single channel quantum wire) at half-filling. The charge 
and spin sectors decouple; the Tomonaga-Luttinger liquid Hamiltonian for the charge sector 
should be supplemented by the Umklapp term which contains only the charge field $\Phi_{\rc}$: 
\bea
{\cal V} = Ua_0(R^+_{\uparrow}L_{\uparrow}R^+_{\downarrow}L_{\downarrow} + h.c.) = 
\frac{U}{2\pi^2 a_0}\cos[\sqrt{8\pi}\Phi_{\rc}],
\eea
where $Ua_0$ is the $2k_F = \pi$ Fourier component of the interaction. The full Hamiltonian
 density for  the charge sector is ${\cal H} = {\cal H}_0 + {\cal V}$, where 
\bea
&&{\cal H}_0 = \frac{1}{2}\left[(\hat \pi_{\rc})^2 + 
(\p_x\Phi_{\rc})\int \rd y V(x-y)(\p_y\Phi_{\rc})\right]\nonumber\\
&&V(x) = v^2\delta(x) + \frac{e^2v}{\pi|x|}\label{H0}
\eea
$v$ being the Fermi velocity. This model is very close to the sine-Gordon one. The spectrum 
contains solitons and their bound states (breathers or excitons). To get their spectrum  one 
can just expand the cosine around its minimum and obtain:
\bea
\omega^2 = (vq)^2 \left[1 + \frac{2e^2}{\pi v}\ln(1/qa_0) \right]
+ m_b^2 , \label{omega}
\eea
where the breather gap is  $
m_b = (2/\pi)\sqrt{U\epsilon_F}, ~~ \epsilon_F = \pi v/a_0$. 
The breathers are effectively optical phonons of the one-dimensional Wigner 
crystal \cite{glazman}. The soliton  gap is larger; it was  estimated in \cite{levitov}, 
a better estimate is 
\bea
M_s = \frac{1}{\pi}[1 + (2e^2/\pi v)\ln(v/M_sa_0)]\sqrt{U\epsilon_F}
\eea
The spectrum in the spin sector is gapless, and the spin velocity is approximately $v$. 
Thus already for a simple one-chain problem, the long-distant Coulomb force brings three 
important new features: (i) strong upward 
renormalization of the charge velocity seen in Eq.(\ref{omega}), and (ii) the presence 
of exciton modes in the charge sector, (iii) power law dependence of the mass scales $M_s, m_b$ on the Umklapp matrix element $U$.  

\section{The excitation spectrum of  nanotube  at half filling}

SWCNs are manufactured by wrapping  two-dimensional graphite sheets into cylinders 
(compactification). The electronic spectrum of an infinite  graphite sheet contains two 
Dirac cones with different chirality, centered at different points of the Brillouin zone 
(the blue dots on  Fig. 1). Under compactification the spectrum is divided into subbands 
corresponding to different quantized values of the transverse momentum. In  the so-called 
armchair nanotubes $k_y =0$ remains an eigevalue and the spectrum in the lowest subband 
stays gapless; at half filling it is also doubly degenerate at low energies. This 
degeneracy is a vestige of the double-cone structure of the two-dimensional dispersion. 
We discuss the spectrum in some detail in Appendix A; more detailed information can be 
obtained from the book \cite{dresselhaus}.  

\begin{figure}[ht]
\begin{center}
\epsfxsize=0.5\textwidth
\epsfbox{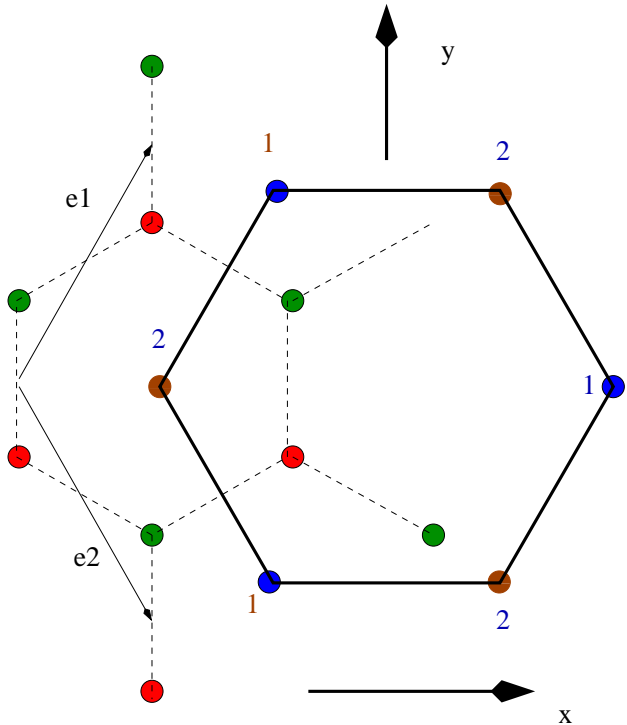}
\end{center}
\caption{The Bravais lattice cell (shown in dotted lines) including two inequivalent carbon 
ions (shown in red and green) and the Brillouin zone for graphite. 
${\bf e}_1$ and ${\bf e}_2$ are basis vectors of the diatomic unit cell.
The positions of the 
tips of two Dirac cones labeled by 1 and 2 are represented by  blue and brown dots.    
\label{zone}} 
\end{figure}

  In Appendix B we derive the bosonized form of the effective low-energy 
Hamiltonian.  The total Hamiltonian 
density is 
\bea
{\cal H} = {\cal H}_0[\Phi_{\rc}^{(+)}] + 
\frac{1}{2}\sum_{a}\left[(\hat \pi^{(a)})^2 + v^2(\p_x\Phi^{(a)})^2\right] + {\cal V} \label{H}
\eea
where ${\cal H}_0$ is given by Eq.(\ref{H0}); $\Phi_{\rc}^{(+)}$ is the symmetric charge 
mode.  The label $a$ takes 
three values, $ a = ({\rc},-), (s,+), (s,-)$, corresponding to the antisymmetric charge 
field and symmetric and antisymmetric spin fields, respectively (see Appendix A).  
The interaction density ${\cal V}$ 
contains Umklapp terms
\bea
&&{\cal V} = -\frac{1}{2(\pi\alpha)^2}\cos(\sqrt{4\pi}\Phi_{\rc}^{(+)})\nonumber\\ 
&& \times 
[g_c\cos(\sqrt{4\pi}\Phi_{\rc}^{(-)}) + (g_3 - g_1)\cos(\sqrt{4\pi}\Phi_s^{(+)}) 
 + g_3\cos(\sqrt{4\pi}\Theta_s^{(-)})  
 - g_1\cos(\sqrt{4\pi}\Phi_s^{(-)})] \nonumber\\
 && + ~\cdots \label{tunn}
\eea
where the dots stand for all other terms that
 do not involve $\Phi_{\rc}^{(+)}$ 
and stay marginal even in the presence of the unscreened Coulomb intreraction.
The couplings $g_c$, $g_{1}$ and $g_3$ are determined by  the lattice Hamiltonian. In Appendix B we 
estimate them for realistic carbon nanotubes, where, as it turns out,  $g_c =g_1$ and, hence, 
there are only two independent 
coupling constants. The unscreened Coulomb interaction strongly reduces the scaling 
dimension of
the operator $\cos(\sqrt{4\pi}\Phi_{\rc}^{(+)})$ in the long-wavelength limit, making it 
smaller than 1.
Thus the Umklapp terms (\ref{tunn}) become strongly relevant, with the scaling dimension 
almost equal to 1.
This circumstance dramatically increases the values of the gaps.

One may well expect that the double degeneracy of the electron band in  carbon nanotubes 
makes the problem similar to the problem of  two interacting chains (the 'ladder' problem) 
much discussed in literature. This is indeed the case. To put the problem in a broader 
context, let us discuss  two extreme types of ladders: (i) the ones where two chains are 
placed far apart such that there is no direct tunneling between them, and (ii) those in 
which the interchain tunneling is stronger than all the interactions.  Hamiltonian 
(\ref{H},\ref{tunn}) is formally equivalent to  the first case describing two well 
separated chains. As a matter of fact, the Hamiltonian for case (ii) is not very 
different. The calculations  done in \cite{fradkin} for the ladder with strong  
interchain tunneling yields the same Hamiltonian as (\ref{tunn}), but with $\Phi_{\rc,-}$ 
field being substituted by its 
dual counterpart, $\Theta_{\rc,-}$:
 \bea
{\cal V}_B = {\cal V}_A[\Phi_{\rc,-}\rightarrow \Theta_{\rc,-}] \label{notunn}
\eea
Therefore one expects that the two models have the same excitation spectrum though the 
response functions are different due to the different field identification. 

 Hamiltonian (\ref{tunn}) is not very convenient to analyse in its bosonic form since 
the effective potential  contains 
mutually nonlocal and
noncommuting fields, $\Phi_{s,-}$ and $\Theta_{s,-}$.  The physics becomes significantly 
more transparent when one uses the refermionization procedure 
\cite{shelton},\cite{shelton2}. Fermionizing  all the fields except of  $\Phi_{\rc}^{(+)}$ 
we obtain
\bea
&&{\cal H} = {\cal H}_0[\Phi_{\rc}^{(+)}] + \ri v(- R^{\dagger}_f \p_x R_f
 + L^{\dagger}_f \p_x L_f)  
+ \frac{\ri v}{2}\sum_{a =0}^3(-\chi_R^a\p_x\chi_R^a + \chi_L^a\p_x\chi_L^a)\nonumber\\
&&-\frac{\ri}{\pi\alpha}\cos(\sqrt{4\pi}\Phi_{\rc}^{(+)})[g_c(R^{\dagger}_f L_f  - h.c.) 
+ 2 g_t \sum_{a =1}^3\chi_R^a\chi_L^a + 2 g_s\chi_R^0\chi_L^0], \label{model}\\
&&~~~g_c = g_1, ~~g_t = (g_3 - g_1), ~~ g_s = - g_1 - g_3,\nonumber 
\eea
where $\chi_R,\chi_L$ are Majorana (real) fermions and $R_f, L_f$ are Dirac fermions 
emerging
from the fermionization of the operator $\cos(\sqrt{4\pi}\Phi_{\rc,-})$ or, in the second case, 
$\cos(\sqrt{4\pi}\Theta_{\rc,-})$. Notice that Eq.(\ref{model}) is manifestly SU(2)-symmetric. 
The triplet of Majorana fermions ($a=1,2,3$)  transform according to the spin 
S=1 representation of the SU(2) group, whereas
the fermion labeled by $a =0$ is a singlet under the SU(2). 
 
 Following \cite{krive}, we shall handle Hamiltonian (\ref{model}) using the adiabatic 
approximation, whose validity is guaranteed by  the fact that the velocity in the 
symmetric charge sector is strongly enhanced  by the long-distant Coulomb 
interaction with respect to the bare Fermi velocity. The results have certain similarity 
with the SU(4) theory  proposed in \cite{balentz}, but also contain important differences, 
which we shall  discuss. 

 Thus, from the $\Phi_{\rc}^{(+)}$-mode point of view, the other degrees of freedom are 
static. Integrating  over this mode, one obtains an  effective potential for the fermions 
in the form of the ground state energy of the sine-Gordon model. According to  
\cite{Zamol} the energy density is proportional to the square of the breather 
mass $\mu_b$ of such sine-Gordon model and in the regime of small sine-Gordon 
coupling constant  $\beta^2$ is equal to
 \bea
E \approx - \frac{\mu_b^2}{4\pi v_c} \approx - v/(\pi\alpha v_c)[ \ri g_c(R^{\dagger}_f L_f - h.c.) + 
2\ri g_t \sum_{a =1}^3\chi_R^a\chi_L^a + 2\ri g_s\chi_R^0\chi_L^0], \label{fast}
\eea
where $v_c \approx v_F[1 + (2e^2/\pi v_F)\ln(\epsilon_F/M)]^{1/2}$ and $v \approx v_F$. 
So as we see, the integration over the fast mode gives rise to the mass terms for all 
of the fermions. The fermionic modes acquire gaps:
\bea
M_{\rc, -} = \frac{v g_c}{\pi \alpha v_c}, ~~ M_{{\rs},t} = \frac{v g_t}{\pi \alpha v_c}, 
~~ M_{{\rs},s} = \frac{v g_s}{\pi \alpha v_c}\label{masses}
\eea
They correspond to neutral excitations with the quantum numbers given in the 
table that follows; see also Fig.2. As far as the fast modes are concerned, one has to 
treat differently the ones 
with and without electric charge. The neutral modes do not involve solitons of field 
$\Phi_{\rc}^{(+)}$. One can get a good estimate of  their spectrum replacing the term 
in the square brackets in expression for the Hamiltonian (\ref{model}) by a constant 
and expanding around the minimum of the cosine potential. As a result one gets the 
same spectrum as for the single chain (\ref{omega}) with  
\bea
m_b^2 = \pi(M_{\rc,-}^2 + 3M_{{\rs},t}^2 + M_{\rs,s}^2)\ln(v/\alpha)\nonumber
 \eea
where with the logarithmic accuracy $M$ is either $M_{\rc,-}$ or $M_S$. 
Thus 
\[
m_b/M \sim [\ln(v/\alpha M)]^{1/2} \gg 1
\]
which further supports the adiabatic approximation. However, as we have already mentioned, 
its validity is already assured by the difference in the velocities. Apart from the breather 
 modes there are  massive modes corresponding to  half-period solitons in $\Phi_{\rc}^{(+)}$ 
with zero modes of the fermions bound  to them. These excitations have  the largest gap 
(we call it $M_e$) and carry  quantum numbers of electron.
 
\smallskip
\begin{center}
Table 1.
\end{center} 
\smallskip
\[
\begin{array}{cccc}\hline\hline
\mbox{Mass} & Q & S & V\\
\hline
M_{{\rs},t}~~ \mbox{triplet} &  0& 1& 0\\
M_{{\rs},s}~~\mbox{singlet} & 0& 0 &0\\
M_{\rc,-} ~~\mbox{vortex}& 0&0& \pm 2\\
m_b ~~\mbox{breathers}& 0&0&0\\
M_e ~~\mbox{electron} &\pm 1&1/2&\pm 1\\
\hline\hline
\end{array}
\]

\begin{figure}[ht]
\begin{center}
\epsfxsize=0.5\textwidth
\epsfbox{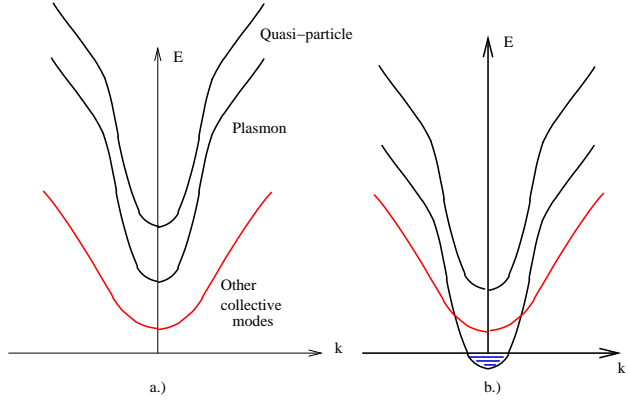}
\end{center}
\caption{A schematic picture of the excitation spectrum: a) at half filling and
b) at small doping. 
\label{spectrum}} 
\end{figure}

The results for the spectrum are summarized in Table 1.;
see also Fig.2.  Excitations are characterized 
by quantum numbers associated with the full continuous symmetry group of the effective model (\ref{model}).
These are the
total charge $Q$ measured in units of the electron charge $e$,  total spin $S$, 
and ``vorticity'' 
$V$. Let us comment on the latter. While the global U(1) phase invariance and the spin SU(2)
symmetry, leading to conservation of the total charge and spin,
\be
Q = \frac{2}{\sqrt{\pi}} \int_{-\infty}^{\infty} \p_x \Phi^{(+)}_{\rc} (x),
~~~S^z = \frac{1}{\sqrt{\pi}} \int_{-\infty}^{\infty} \p_x \Phi^{(+)}_{s} (x),
\label{top-charge}
\ee
are exact symmetries of the original, microscopic
Hamiltonian, an extra (``flavor'') U(1) symmetry generated by global phase transformations of 
the $f$-spinor $(R_f, L_f)$
(or equivalently, by uniform translations of the dual field $\Theta^{(-)}_c$),
emerges in the low-energy limit only.
 This symmetry leads to conservation
of the ``flavor'' charge
\be
V = \frac{2}{\sqrt{\pi}} \int_{-\infty}^{\infty} \p_x \Phi^{(-)}_{\rc} (x),
\label{top-flavor}
\ee
which, together with conservation of $Q$,  
implies independent conservation of the particle numbers
at each Dirac point, $Q_{1,2} = (Q \pm V)/2$.

\begin{figure}[ht]
\begin{center}
\epsfxsize=0.5\textwidth
\epsfbox{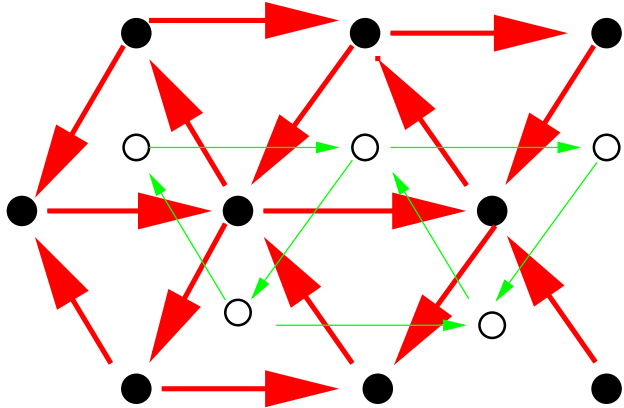}
\end{center}
\caption{A staggered flux (orbital antiferromagnet) state corresponding to a nonzero value
of vorticity $V$. The arrows indicate local currents flowing across the links of the
$a$ (red) and $b$ (green) sublattices.
\label{flux}} 
\end{figure}

The nomenclature ``vorticity'' we have chosen follows from the microscopic origin
of the flavor density. Indeed, consider a lattice operator describing a current flowing around
the elementary plaquette of sublattice $a$ or $b$:
\bea
&& j_{{\rm plaq}, \nu} ({\bf r})
= \ri \left[ \psi^{\dagger}_{\nu}({\bf r}) \psi_{\nu}({\bf r} - {\bf e}_2)
+ \psi^{\dagger} _{\nu}({\bf r} - {\bf e}_2) \psi_{\nu} ({\bf r} + {\bf e}_1)
+ \psi^{\dagger} _{\nu} ({\bf r} + {\bf e}_1) \psi_{\nu} ({\bf r}) - h.c.\right]
\label{curr} \\
&& \nu = a,b \nonumber
\eea
From this construction it follows that the state with a nonzero $\la j_{\rm plaq, a(b)} \ra$
would represent an orbital antiferromanet, or a staggered flux phase, realized on the sublattice
$a$ ($b$), as shown in Fig.\ref{flux}.
Projecting (\ref{curr}) onto the $k_y = 0$ subband and passing to the continuum limit,
one makes sure that the sum ${\bf j}_{{\rm plaq}, a} + {\bf j}_{{\rm plaq}, b}$
does transform to the flavor charge density
\[
\rho_f = \sum_{\s} \left[ :R^{\dagger}_{1\s} R_{1\s}: -  :R^{\dagger}_{2\s} R_{2\s}:
+ ~(R \to L) \right] \sim \p_x \Phi^{(-)}_{\rc}.
\]


Previous attempts to study the problem of two 
coupled channels have mostly relied on the assumption of equal velocities 
with the subsequent use of  RG analysis.  The   approach was
 pioneered by Lin {\it et al.} \cite{balentz} who argued that at strong coupling the 
spectrum of the two-chain problem at half filling  acquires a higher symmetry, such as SO(6) 
or even SO(8). 
In fact,  Gell-Mann-Low equations alone are insufficient to extract information 
about  strong coupling regime (they have to be supplemented by Callan-Symanzik equations 
for the physical observables) and therefore  cannot provide a legitimate  ground for  
such conclusions. Our approach is based on a different assumption; here  the long-range 
Coulomb interaction 
legitimates a clear  separation of scales between single particle excitations and 
collective modes. This  significantly simplifies the calculations providing  one with 
the  well-controlled approximation. It is instructive to compare the  results with 
those conjectured in \cite{balentz}. 
Though the  structure of the multiplets is the same as in  the SU(4)$\sim$SO(6) theory, 
there are important differences in the spectrum. Two  quasiparticle multiplets (particles 
and antiparticles) are four-fold degenerate, as in the SU(4) theory, but  the six-fold 
degenerate multiplet of the SU(4) is split into a doublet with the mass $M_{\rc,-}$, a 
magnetic triplet  with the mass $M_{{\rs},t}$ and a singlet mode with the mass 
$M_{{\rs},s}$. The ratios of  the quasiparticle mass to the masses of 
 the neutral 
modes are vastly different from the SU(4) Gross-Neveu model ratio 
$M_{{\rs}}/M_e = \sqrt 2$. They depend on the Coulomb interaction and the 
quasiparticle gap is much larger than the spin  and the parity collective mode gaps.
The gaps of the collective modes are not equal. In nanotubes where there are only two 
independent coupling constants we expect that 
\bea
2M_{\rc} + M_t + M_s = 0
\eea
(recall that since we deal with Majorana fermions, the masses may be negative,  the spectral gaps being  their 
 absolute values). 

\section{The order parameters of different phases}

 From now on we shall concentrate on the model describing carbon nanotubes.  Different 
phases, shown in Fig. 3, correspond to different ground state phase lockings and are determined by the 
signs of the fermionic masses\cite{shelton}, Eqs.(\ref{masses}). 
These sign 
changes are not reflected in the thermodynamics which is sensitive only to absolute 
values of the masses. The difference in the correlation functions may, however, be 
quite dramatic. 

We use the following conventions \cite{shelton}:
\bea
&&\cos(\sqrt\pi\Phi_s^{(+)}) = \s_1\s_2, ~~ \sin(\sqrt\pi\Phi_s^{(+)}) = \mu_1\mu_2\nonumber\\
&&\cos(\sqrt\pi\Theta_s^{(+)}) = \mu_1\s_2, ~~ \sin(\sqrt\pi\Theta_s^{(+)}) = \s_1\mu_2 \\
&&\cos(\sqrt\pi\Phi_s^{(-)}) = \s_0\s_3, ~~ \sin(\sqrt\pi\Phi_s^{(-)}) = \mu_0\mu_3\nonumber\\
&&\cos(\sqrt\pi\Theta_s^{(-)}) = \mu_0\s_3, ~~ \sin(\sqrt\pi\Theta_s^{(-)}) = \s_0\mu_3 
\eea
where $\{ \s_i \}$ and $\{ \mu_i\}$ $(i=1,2,3,0)$ are order and disorder parameters
of the 2D Ising models associated with the singlet ($i=1,2,3$) and triplet $(i=0)$
Majorana fermions. A particular Ising model is ordered ($\la \s \ra \neq 0$)
or disordered ($\la \mu \ra \neq 0$) depending on the sign of the corresponding Majorana
mass, $M <0$ or $M>0$.
In order to understand the structure and properties of the correlation functions, one 
has to recall that in the ordered phase of the Ising model, where $\la\s\ra \neq 0$, 
the correlation function $\la\la\mu(\omega,q)\mu(-\omega,-q)\ra\ra$ contains a coherent 
peak, while the correlation function of the $\s$'s does not.  

There are  six possible phases, two  of them being Haldane spin liquids. Such liquids are characterized by the presence of  a coherent triplet magnetic excitation (magnon) in the  two-point  correlation function
of the staggered magnetizations. 
In non-Haldane 
disordered phases,
the spectrum of triplet excitations is incoherent; however spin-singlet modes may be coherent.
Table 2 and Fig. 3 show what 
operators acquire nonzero expectation values in the corresponding ground  states. 
It will be assumed that in all phases
the symmetric charge field is locked at $\Phi^{(+)}_{\rc} = 0$, so that
$\la \cos \sqrt{\pi} \Phi^{(+)}_{\rc} \ra \neq 0$,
$\la \sin \sqrt{\pi} \Phi^{(+)}_{\rc} \ra = 0$.

\newpage

\begin{center}
Table 2.
\end{center}
\smallskip
\[
\begin{array}{cccc}\hline\hline
\mbox{Phase} &  & &  \\
\hline
A &  \la\cos(\sqrt\pi\Phi_{\rc}^{(-)})\ra \neq 0& \la\s_a\ra \neq 0& \la\s_0\ra \neq 0\\
\hline
B& \la\cos(\sqrt\pi\Phi_{\rc}^{(-)})\ra \neq 0& \la\mu_a\ra \neq 0 &\la\s_0\ra \neq 0\\

\hline
C& \la\sin(\sqrt\pi\Phi_{\rc}^{(-)})\ra \neq 0&\la\mu_a\ra \neq 0 & \la\s_0\ra \neq 0\\
\hline
D& \la\sin(\sqrt\pi\Phi_{\rc}^{(-)})\ra \neq 0& \la\mu_a\ra \neq 0& \la \mu_0 \ra \neq 0\\
\hline
 E&\la\sin(\sqrt\pi\Phi_{\rc}^{(-)})\ra \neq 0&\la\s_a\ra \neq 0 &\la \mu_0 \ra \neq 0
\\
\hline
F &\la\cos(\sqrt\pi\Phi_{\rc}^{(-)})\ra \neq 0 &\la\s_a\ra \neq 0 &\la \mu_0 \ra \neq 0\\
\hline\hline
\end{array}
\]

\begin{figure}[ht]
\begin{center}
\epsfxsize=0.5\textwidth
\epsfbox{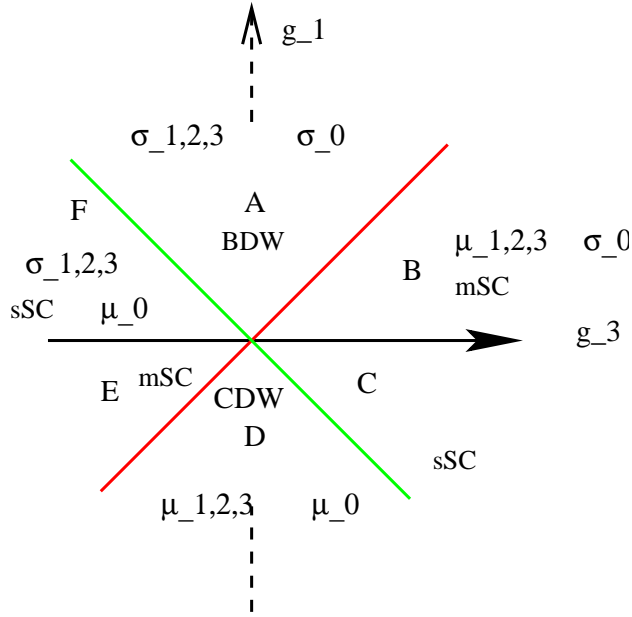}
\end{center}
\caption{The phase diagram. For each quadrant  it is indicated which Ising models order and 
disorder parameters have non-zero ground state expectation values. The phases are separated by critical 
lines on which one of the particle masses vanishes. The green and red lines are $Z_2$ and $SU_2(2)$ critical 
lines respectively. The $g_1 =0$ axis corresponds to the $U(1)$ critical line. Phases A and D have density wave order. Under doping phases B, E and C,F develop  a power law response to superconducting paring.  
\label{phase}} 
\end{figure}
Below we give a brief characterization of each phase;  for more details see 
the subsequent discussion. 
\begin{itemize}
\item
Phase $A$: $g_1 > |g_3|$. This phase has a long-range Bond Density Wave (BDW)
order at $T =0$ (see Fig. 4 below).
\item
Phase $B$: $0< g_1 < g_3$. This is a Haldane spin-liquid phase whose excitation spectrum contains
a coherent triplet 
magnon with the mass $M_{S,t}$, associated with the correlation function of the 
site-diagonal spin operator
${\bf S}^{(-)}$; see Eq. (\ref{S-:trip}). 
Under doping it 
develops a power-law response of the pairing susceptibility of type described further in the text. The corresponding superconducting order parameter transforms non-trivially under the lattice point group.
\item
Non-Haldane phase $C$: $- g_3 < g_1 <0$. The magnetic 
singlet mode with mass $M_{S,s}$ becomes coherent in this phase. As a result a  coherent peak  appears in the correlation function of 
the Charge Density Wave order parameter $\Delta_{CDW}$, Eq. (\ref{CDW-op}). Under doping this phase develops superconducting correlations in the s-channel. 
\item
Phase $D$: $ g_1 < - |g_3|$. There is a Charge Density Wave (CDW) long-range order in the
ground state of this phase. Phases CDW and BDW are mutually dual.
\item
Phase $E$: $g_3 < g_1 < 0$. This is another Haldane phase dual to phase $B$.
Phase E has a coherent magnon in the site-off-diagonal spin density  $\vSigma^{(-)}$, Eq.(\ref{Sigma-op}).
Under doping it develops superconducting correlations of the same type as phase B. 
\item
Phase $F$: $0 < g_1 < -g_3$. This phase
is dual to phase $C$ and has a coherent spin-singlet mode displayed by the BDW order-parameter correlation function. Under doping it develops superconducting correlations in the s-channel. 
\end{itemize}

\begin{figure}[ht]
\begin{center}
\epsfxsize=0.5\textwidth
\epsfbox{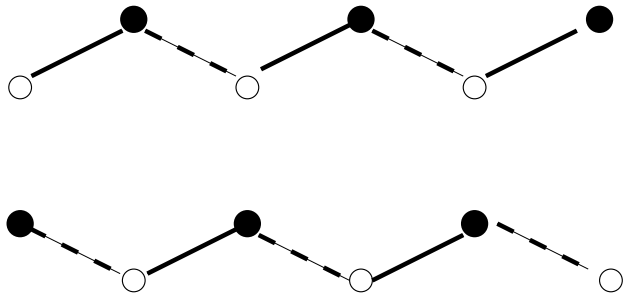}
\end{center}
\caption{The dimerization in BDW phase    
\label{bdw}} 
\end{figure}

\subsection{Order parameters, possible orderings and dominant correlations}

{\bf CDW}. Already for 
 geometrical reasons (the lattice is bipartite) one may expect that
the system at half filling can develop a commensurate Charge Density Wave.
Introducing locally averaged electron densities for $a$ and $b$ sublattices,
\be
\bar{\rho}_{\nu} ({\bf r})
= \frac{1}{3} \left[
\psi^{\dagger}_{\nu} \psi_{\nu}({\bf r}) + \psi^{\dagger}_{\nu} \psi_{\nu}({\bf r}+ {\bf e}_1)
+ \psi^{\dagger}_{\nu} \psi_{\nu}({\bf r}- {\bf e}_2)
\right], ~~~\nu = a,b
\label{loc-ave}
\ee
we define the
CDW order parameter (OP) as follows:
\bea
&&\Delta_{CDW}({\bf r}) = \bar{\rho}_{a} ({\bf r}) - \bar{\rho}_{b} ({\bf r}) 
\nn
&&~~  \sim \sum_{\s} \left( R^{\dagger}_{1\s} L_{1\s} -
R^{\dagger}_{2\s} L_{2\s} \right) + h.c.\nn
&&~~
\sim  \sin(\sqrt{\pi}\Phi_{\rc}^{(+)})\cos(\sqrt\pi\Phi_{\rc}^{(-)})\s_1\s_2\s_3\s_0 - 
\cos(\sqrt{\pi}\Phi_{\rc}^{(+)})\sin(\sqrt\pi\Phi_{\rc}^{(-)})\mu_1\mu_2\mu_3\mu_0
\label{CDW-op}
\eea
This OP has a non-zero average value in phase $D$. In phase $C$ 
where $\la\sin(\sqrt\pi\Phi_{\rc}^{(-)})\ra \neq 0$ and $\la\mu_a\ra \neq 0$ (a =1,2,3), the most singular 
part of the CDW order parameter is  proportional to $\mu_0$. Since we are in the phase with $\la\s_0\ra \neq 0$, 
operator $\mu_0$ has a non-zero matrix element between the ground state and a state with one Majorana fermion. 
Therefore the correlation 
function of CDW OPs contains a coherent peak corresponding to an emission of the singlet 
magnetic mode with the mass $M_{S,s}$. 

{\bf BDW}. The Bond Density Wave order parameter is similar to the CDW one, but is off-diagonal in the site indices:
\bea
&&\Delta_{BDW}({\bf r}) = \psi^{\dagger}_a ({\bf r})[\psi_b({\bf r} + {\bf e}_1)- \psi_b({\bf r} - {\bf e}_2)]  
- [\psi^{\dagger} _a({\bf r} + {\bf e}_1)- \psi^{\dagger} _a ({\bf r} - {\bf e}_2)] \psi_b({\bf r}) +
h.c.
\nn
&&~~ \sim 
\ri\sum_{\s}\left( R^{\dagger}_{1\s} L_{1\s} - R^{\dagger}_{2\s} L_{2\s} \right) + h.c.  \nonumber\\
&& \sim \sin(\sqrt{\pi}\Phi_{\rc}^{(+)})\sin(\sqrt\pi\Phi_{\rc}^{(-)})\mu_1\mu_2\mu_3\mu_0 + 
\cos(\sqrt{\pi}\Phi_{\rc}^{(+)})\cos(\sqrt\pi\Phi_{\rc}^{(-)})\s_1\s_2\s_3\s_0.
\label{bdw-op}
\eea
This phase is dual to the CDW one.
The BDW order parameter condenses in phase A. 
The dimerization ordering pattern in this phase is shown in Fig. 4.
In the disordered phase F (which is dual to phase C), the spectral weight of the OP 
$\Delta_{BDW}$ contains a coherent magnetic singlet mode with the mass 
$M_{S,s}$. 

Notice that the CDW and BDW OPs do not contain any oscillatory pieces. 
These show up in the density distributions that are not uniform
across a sublattice but otherwise are consistent with the uniaxial symmetry
of the nanotube:
\[
\tilde{\rho}_{\nu} ({\bf r}) =
\frac{1}{2} \left[ \psi^{\dagger}_{\nu} \psi_{\nu}({\bf r}+ {\bf e}_1)
+ \psi^{\dagger}_{\nu} \psi_{\nu}({\bf r}- {\bf e}_2) -
 \psi^{\dagger}_{\nu} \psi_{\nu}({\bf r})\right], ~~~\nu = a,b.
\]
Two 
incommensurate 
CDW OPs can then be constructed as follows:
\bea
\Delta^{(+)}_{mCDW}({\bf r}) &=& \tilde{\rho}_{a} ({\bf r}) + \tilde{\rho}_{b} ({\bf r}) \nn
&\sim& \re^{- 2iQx} \sum_{\s} \left( L^{\dagger}_{1\s} R_{2\s} - R^{\dagger}_{1\s} L_{2\s} \right) + h.c.\sim  \ri \kappa_{1\up} \kappa_{2\up} \re^{- 2iQx} \re^{i \sqrt{\pi} \Theta_c ^{(-)}}\nn
&\times&
\left[ \sin (\sqrt{\pi} \Phi_c ^{(+)}) \mu_1 \mu_2  \mu_3 \s_0 +
\ri \cos (\sqrt{\pi} \Phi^{(+)}) \s_1 \s_2 \s_3 \mu_0\right]
 + h.c.;
\label{mCDW+} \\
\Delta^{(-)}_{mCDW}({\bf r}) &=& \tilde{\rho}_{a} ({\bf r}) - \tilde{\rho}_{b} ({\bf r}) \nn
&\sim& \re^{- 2iQx}  \sum_{\s} \left( R^{\dagger}_{1\s} R_{2\s} - L^{\dagger}_{1\s} L_{2\s}\right) + h.c.
\nn
&\sim& \kappa_{1\up} \kappa_{2\up}\re^{-2\ri Q x}
[R^{\dagger}_f \chi_R^{(0)} + L^{\dagger}_f \chi_L^{(0)}] + h.c.
\label{mCDW-}
\eea
where $\kappa$'s are Klein factors. Since the dual  antisymmetric charge field, $\Theta^{(-)}_c$, is strongly
disordered ($\la \re^{i \sqrt{\pi} \Theta_c ^{(-)}} \ra = 0$), the OP $\Delta^{(+)}_{mCDW}$
has zero expectation value in all sectors of the phase diagram. In phase F, where
$\la \s_{a} \ra \neq 0 ~(a=1,2,3)$, $\la \mu_0 \ra \neq 0$, the correlation function of
the mCDW$^+$ OP displays a coherent peak corresponding to emission of a vortex particle with the mass $M_{c,-}$.

\begin{figure}[ht]
\begin{center}
\epsfxsize=0.5\textwidth
\epsfbox{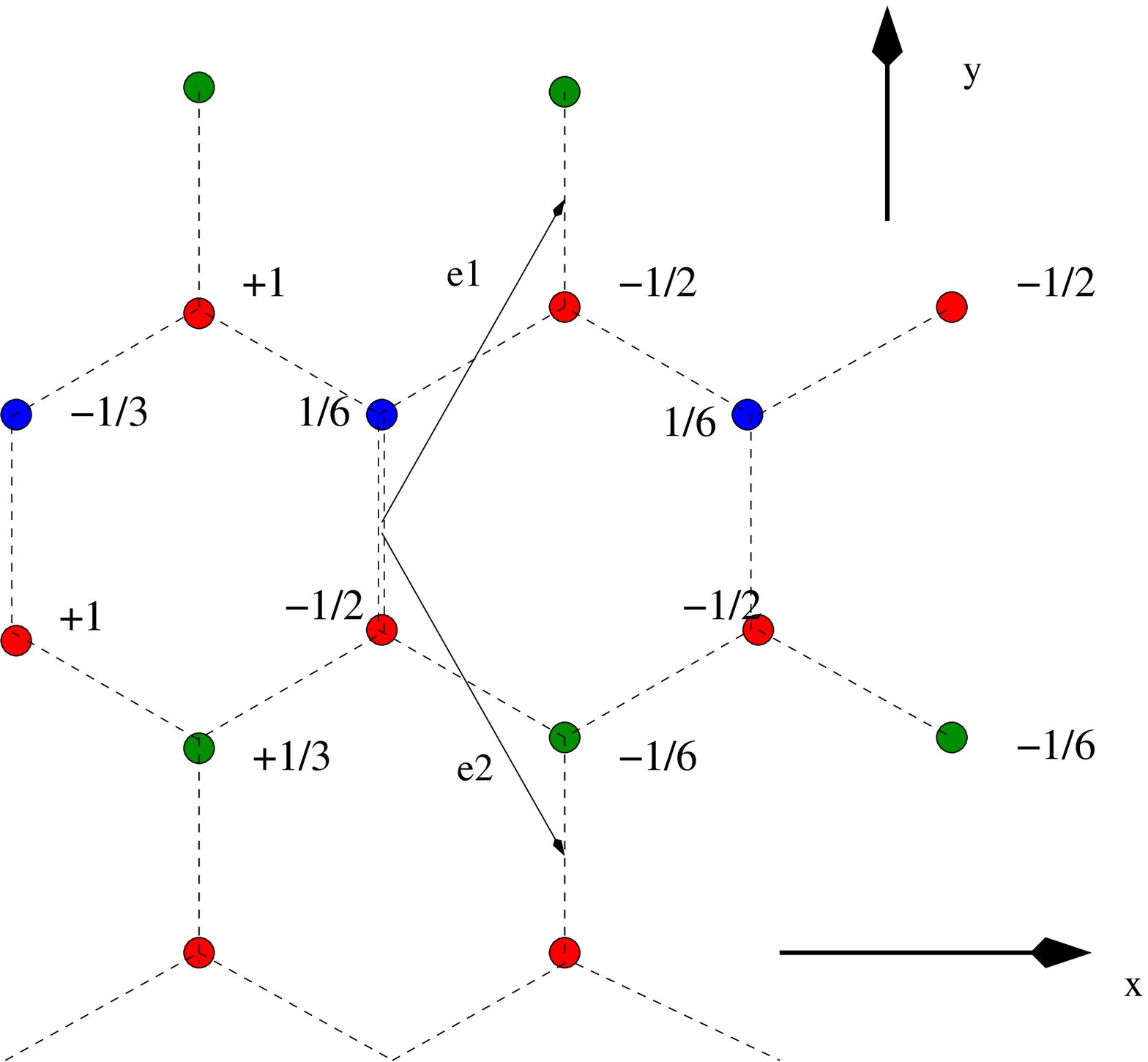}
\end{center}
\caption{The charge distribution for the mCDW order.  
\label{cdw}} 
\end{figure}

The charge distribution corresponding to the OP $\Delta^{(-)}_{mCDW}$
is depicted on Fig. 
\ref{cdw}. This kind of order can be induced  by applying a modulated potential of 
sufficient strength that couples to $\Delta_{mCDW}^-$. The potential must be strong enough 
to  overcome the energy gaps of the corresponding excitation branches and to drive 
the system into a  state with an induced order of the mCDW$^-$ type.

{\bf Magnetization}.  The total magnetization is given by 
\bea
{\bf S}_+ &=& \frac{1}{2} \psi^{\dagger}_a\vsigma \psi_a + \frac{1}{2}\psi^{\dagger}_b\vsigma \psi_b  \nonumber\\
&=& {\bf I}_R + {\bf I}_L +
\frac{1}{2} \{\re^{2 i Q x}[(R^+_2 \vsigma L_1) - (L^+_2\vsigma R_1)] + h.c.\}\nonumber
\eea
Its smooth part is equal to the sum of  the chiral currents of the Majorana triplet:
\[
I^a _{R(L)} = J^a _{1,R(L)} + J^a _{2,R(L)} = - \frac{\ri}{2} \eps^{abc}\chi^b _{R(L)}\chi^c _{R(L)}.
\] 
The oscillating part of the spin density is (Klein factors omitted)
\bea
{\bf S}^{(+)}_{2Q} \sim \re^{ \ri\sqrt\pi\Theta_{\rc}^{(-)}}
\left[ - \sin(\sqrt\pi\Phi_{\rc}^{(+)})\mu_0{\bf N} + \ri\cos(\sqrt\pi\Phi_{\rc}^{(+)})
\s_0\tilde{\bf N}\right].
\eea
Here $\tilde{\bf N} = ( \mu_1\s_2\s_3, \s_1\mu_2\s_3, \s_1\s_2\mu_3)$ and 
${\bf N} = (\s_1\mu_2\s_3 , \mu_1\s_2\mu_3,\mu_1\mu_2\s_3 )$. 
This OP  is never coherent.

On the other hand, to define the {\bf staggered magnetization} 
${\bf S}^{(-)} ({\bf r})$
associated with the bipartite 
lattice, we need to introduce locally averaged spin densities for the two sublattices
(cf. Eq. (\ref{loc-ave})):
\be
\bar{\bf s}_{\nu} ({\bf r}) = \frac{1}{6} \left[ \psi^{\dagger}_{\nu} \vsigma \psi_{\nu}({\bf r})
+ \psi^{\dagger}_{\nu} \vsigma \psi_{\nu}({\bf r}+{\bf e}_1)
+ \psi^{\dagger}_{\nu} \vsigma \psi_{\nu}({\bf r}-{\bf e}_2)
\right], ~~~\nu = a,b
\label{loc-ave-spin}
\ee
Then 
\bea
{\bf S}^{(-)}({\bf r}) &=& \bar{\bf s}_{a} ({\bf r}) - \bar{\bf s}_{b} ({\bf r})\nn
&\sim& \left( R^{\dagger}_1 \vsigma L_1 - R^{\dagger}_2 \vsigma L_2 \right) + h.c. \nn
&\sim& \cos(\sqrt\pi\Phi_{\rc}^{(+)})\cos(\sqrt\pi\Phi_{\rc}^{(-)})\s_0{\bf N} - 
\sin(\sqrt\pi\Phi_{\rc}^{(+)})\sin(\sqrt\pi\Phi_{\rc}^{(-)})\mu_0\tilde{\bf N} \label{S-:trip}
\eea
The OP ${\bf S}^{(-)} ({\bf r})$ is  coherent in the Haldane spin-liquid phase $B$, the corresponding particle
representing a massive triplet magnon. Notice that in the expression (\ref{S-:trip})
the vector field ${\bf N}$ plays the role of the staggered magnetization of the effective
antiferromagnetic spin-1 chain \cite{shelton}.

In full analogy with the BDW OP (\ref{bdw-op}), one can construct the site-off-diagonal staggered
magnetization:
\bea
\vSigma^{(-)} (\bf r) &=&
\frac{1}{2} \psi^{\dagger}_a ({\bf r}) \vsigma \left[
\psi_b ({\bf r}+{\bf e}_1) - \psi_b ({\bf r}-{\bf e}_2)
\right]
- \frac{1}{2} \left[
\psi^{\dagger}_a ({\bf r}+{\bf e}_1) - \psi^{\dagger}_a ({\bf r}-{\bf e}_2)
\right] \s \psi_b ({\bf r}) \nn
&\sim& \frac{\ri}{2} \left( R^{\dagger}_1 \vsigma L_1 - R^{\dagger}_2 \vsigma L_2 \right) + h.c.
\nn
&\sim& \cos \sqrt{\pi} \Phi^{(+)}_{\rc} \sin \sqrt{\pi} \Phi^{(-)}_{\rc}
\mu_0 {\bf N} + 
 \sin\sqrt{\pi} \Phi^{(+)}_{\rc} \cos \sqrt{\pi} \Phi^{(-)}_{\rc}
\s_0 \tilde{\bf N}
\label{Sigma-op}
\eea
The structure of (\ref{Sigma-op}) indicates that phase $E$ represents a Haldane spin liquid
and is dual to phase $B$. Indeed, in 
phase $E$ it is the vector field $\tilde{\bf N}$ that can be regarded as the staggered magnetization
of the effective S=1 chain, and, hence, the spectral weight of the operator $\vSigma^{(-)}$
contains in this phase a coherent triplet magnon.

{\bf Pairing operators}. Let us construct the {\bf s-wave superconducting (SC) order parameter}
$\Delta_{sSC}$.
To this end, we first build locally averaged, site-diagonal, singlet pairing operators:
\[
\Delta^{\nu}_{sSC} ({\bf r}) =
\frac{1}{3} \left[
\psi_{\nu, \up} \psi_{\nu, \down} ({\bf r}) + \psi_{\nu, \up} \psi_{\nu, \down}
({\bf r} + {\bf e}_1) + \psi_{\nu, \up} \psi_{\nu, \down} ({\bf r} - {\bf e}_2)
\right], ~~~\nu = a,b.
\]
The OP $\Delta_{sSC}$ is then defined as
\bea
\Delta_{sSC} ({\bf r}) &=&  \Delta^a _{sSC} ({\bf r}) + \Delta^b _{sSC} ({\bf r})\nn
&\sim& \sum_{\s} \s \left( L_{1\s} R_{2, -\s} -  L_{2\s} R_{1, -\s}\right) \nn
&\sim& \kappa_{1\up} \kappa_{2\down} e^{-i \sqrt{\pi} \Theta^{(+)}_{\rc}}
\left( \cos \sqrt{\pi} \Phi^{(-)}_{\rc} 
\s_1 \s_2 \s_3 \mu_0
+ \ri \sin \sqrt{\pi} \Phi^{(-)}_{\rc} 
\mu_1 \mu_2 \mu_3 \s_0 \right).
\label{sSC-op}
\eea
The amplitude of this OP acquires a finite value in non-Haldane phases $C$ and $F$. These
two phases, with coherent singlet magnetic modes in their spectrum at half-filling,
exhibit power-law sSC correlations under doping.


 The s-wave type superconductivity is not the only possible singlet SC order one can 
imagine. One can introduce  {\bf SC order} parameter which transforms non-trivially 
under the point group: 
\bea
\Delta_{mSC} &=& \frac{1}{3}\psi_{a\up}({\bf r})[\psi_{b\down}({\bf r}) - 
\psi_{b\down}({\bf r} +{\bf e}_1) - \psi_{b\down}({\bf r} -{\bf e}_2)] - 
(\up \leftrightarrow \down) \nn
&\sim& \sum_{\s} \s \left( L_{1\s} R_{2, -\s} +  L_{2\s} R_{1, -\s} \right)
+ {\rm oscillatory~ terms}
\nn
&\sim& \re^{-i \sqrt\pi\Theta_{\rc}^{(+)}}\left[\cos(\sqrt\pi\Phi_{\rc}^{(-)})
\mu_1\mu_2\mu_3\s_0
+ \ri\sin(\sqrt\pi\Phi_{\rc}^{(-)})\s_1\s_2\s_3\mu_0 \right].
\label{mSC-op}
\eea
This pairing OP exhibits power-law correlations in doped Haldane phases
$B$ and $E$. 

 To see what kind of superconductivity this is, let us consider the lattice mean 
field Hamiltonian:
\bea
H = (\psi^+_{a\up},\psi^+_{b\down},\psi_{a\down},\psi_{b\up})\left(
\begin{array}{cccc}
0 & t(k) & 0 &g(k)\Delta\\
t^*(k) & 0& -g(k)\Delta & 0\\
0 & - g^*(k)\Delta^* & 0 & t(k)\\
g^*(k)\Delta^* & 0 &t^*
(k) &0
\end{array}
\right)
\left(
\begin{array}{c}
\psi_{a\up}\\
\psi_{b-\down}\\
\psi^+_{a\down}\\
\psi^+_{b-\up}
\end{array}
\right)
\eea
where $g(k) = -1 + 2\cos(k_x/2)\re^{\ri\sqrt 3 k_y/2}$. The spectrum is 
\bea
E^2 = |t(k)|^2 + |g(k)|^2|\Delta|^2 .
\eea

In the Haldane phase $
C$ under doping we get the following OP:
\bea
\psi_{a\s}\p_x\psi_{b-\s} - \psi_{a-\s}\p_x\psi_{b\s}
\eea

The {\bf Aharonov-Bohm} effect: the $y$-component of the vector potential is coupled to
\bea
&&(R^+_{1\s}L_{1\s} + L^+_{1\s}R_{1\s}) - (R^+_{2\s}L_{2\s} + L^+_{2\s}R_{2\s}) 
\sim\nonumber\\
&&\cos(\sqrt\pi\Phi_{\rc}^{(+)})\sin(\sqrt\pi\Phi_{\rc}^{(-)})\s_1\s_2\s_3\s_0 - 
\sin(\sqrt\pi\Phi_{\rc}^{(+)})\cos(\sqrt\pi\Phi_{\rc}^{(-)})\mu_1\mu_2\mu_3\mu_0
\eea
It is coherent in Haldane phase $E$ with an emission of the magnetic singlet.

The {\bf staggered flux} around a hexagonal plaquette is
\bea
\Phi = 6\re^{\ri\pi/6}\re^{2\ri Qx}(L^+_2L_1 - R^+_2R_1) + h.c. \sim 
(R_f\chi_R^{(0)} - L_f\chi_L^{(0)})\re^{2\ri Qx + \ri\pi/6} + h.c.
\eea
where the flavor fermion is associated with $\Phi_{\rc}^{(-)}$ field. Therefore 
there is an interesting possibility 
to drive a system into a peculiar critical state with a mixture of the magnetic 
singlet and nonmagnetic orbital mode by applying a magnetic field with the corresponding period. 

\section{Small doping}

  The large velocity difference between the symmetric charge modes and the other part 
of the spectrum holds a key to the stability of the approach  at finite doping. For 
small doping $k_Fa \ll \exp(- \pi v/2e^2)$ the soliton mode of  $\Phi_{\rc}^{(+)}$ 
field, though becoming gapless,   
still lies above the others in the most of the momentum space (see Fig. 2b). This means 
that one can still integrate over $\Phi_{\rc}^{(+)}$ and obtain  Eq.(\ref{fast}) though  
with a prefactor $< 1$. Thus a small doping will decrease the gaps. It may also give rise 
to finite decay rates for the collective modes. This is a pseudogap regime which is perhaps 
similar to the one existing in the underdoped state of the cuprate superconductors (see, 
for example, the recent paper \cite{sudip}). The physics of this regime will be goverened 
by two energy scales: the scale of collective modes $M_c$ (see Eq.(\ref{masses}) and the 
electronic scale $M_e$. At energies smaller than $M_c$ the system is a Luttinger liquid, at 
$M_c < E < M_e$ the collective modes contribute to all physical properties giving rise to 
strong enhancement of the magnetic susceptibility and specific heat. Above  $M_e$ the 
effects of backscattering disappear. 

 Let us briefly discuss a possible  enhancement of the superconducting fluctuations in a 
doped regime. In those phases where the amplitude of the SC order parameter is frozen 
(see the discussion above) its power law correlations are determined by the scaling 
dimension of operator $\exp[\ri\sqrt\pi\Theta_{+}^{(c)}]$. This scaling dimension is 
equal to $1/4\tilde K_{\rc}$ where $\tilde K_{\rc}$ is the renormalized  Luttinger constant. 
At large doping the value of this constant  is determined by the 
long-distant Coulomb interaction and is small, but close to the transition (small doping) 
it is close to one (see \cite{papa} for a more detailed analysis). Therefore there is a 
window of doping where $\tilde K_{\rc} < 1/2$ where the pairing susceptibility diverges.  
A more detailed description of the pseudogap regime will be given elsewhere.   

\section{Conclusion}

 Let us summarize our results. We have 
analyzed the effect of the long-range Coulomb interaction on the low-energy properties
of the armchain nanotube. We have shown that the spectrum of the system is
very rich. The phase diagram includes six  phases differing  from each other by their 
response functions. 
Under doping four of them have enchanced superconducting fluctuations. 
\bigskip

SWCNs exhibit two types of behavior: either band insulators or Luttinger liquids.
Armchair CNs are believed to be metallic and be of the Tomonaga-Luttinger (TLL) type. This, in fact, is
only true away from 1/2 filling. As shown in this paper, the picture changes dramatically
at 1/2 filling when effects of strong correlations become crucial. To observe these effects
in experimental conditions,
the chemical potential must be fine tuned to the value $\mu = 0$ at which the Fermi ``surface''
is represented by two degeneracy (Dirac) points. Then, due to the special role
of Umklapp processes in the presence of unscreen Coulomb interaction,
on lowering the temperature,
one must be able to see a crossover from the metallic, TLL behavior to an insulator
discussed in the paper. Such insulating state is Mott-type, with a much richer spectrum
than in the Hubbard model. One important difference is that, unlike in the Hubbard model, the spin excitations have spectral gaps. 

 The estimates of interaction matrix elements for carbon nanotubes are provided in \cite{odintsov}:
\bea
&&g_i = \beta_i \frac{e^2}{N\epsilon}, ~~ i = c,t,s \nn
&& \beta_c \approx 0.4, ~~\beta_t \approx 0.5, ~~\beta_s \approx - 1.3
\eea
where $N$ is the number of transverse bands of the nanotube and $\epsilon \approx 1.4$ is the dielectric constant. This estimates indicate that realistic nanotubes belong to phase B. Eq.(\ref{masses}) also  gives the following estimates for the spectral gaps of the collective modes:
\bea
|M_i| = |\beta_i| \frac{e^2}{\pi N\alpha\epsilon}\frac{v}{v_c}
\eea
Since $\alpha$ is the short distance cut-off of the bosonic theory and is therefore non universal, this formula contains a certain degree of ambiguity. For noninteracting electrons $\pi\alpha = a_0$ (the lattice spacing). Taking $N =10$, $v/v_c =1/3$ and $a_0 = 0.246$nm we get the estimate of order of 0.1 ev. The single particle gaps should be even greater. 

 Since the gaps should be  quite sizeable, one may wonder whether they have not been already observed. The existing techniques produce simultaneously carbon nanotubes of different sizes and chiralities and one has to select the relevant ones using some criteria. The simplest one is to differ between metallic and insulating (semiconductor) nanotubes. Naturally, on the first glance Mott and band insulators look alike. Therefore it is possible  that the Mott insulating armchair nanotubes 
have been overlooked being taken for semiconducting  ones. To distinguish Mott  insulator from a band one one has to measure transport and magnetic properties and compare the gaps. The clearest sign of 'mottness' is the difference in gap sizes in different response functions. This is the feature to look for experimentally.

\acknowledgements

 We are    grateful to P. Azaria, A.O. Gogolin,
 Ph. Lecheminant, A. Chubukov, P. Coleman and  F. H. L. Essler for fruitful discussions.  
AMT  acknowledges a  support from US DOE under contract No. DE-AC02-98CH10886 and a kind 
hospitality of Abdus Salam ICTP where a part of the work was conducted. 
AAN is partly supported by MIUR, under project COFIN2003 ``Field Theory, Statistical 
Mechanics and Electron Systems''. He also 
acknowledges the support from the Institute for Theory of Strongly Correlated and Complex Systems
at Brookhaven. 
 
\appendix 
 
\section{}

 The original Hamiltonian describing noninteracting electrons on a 
honeycomb lattice can be compactly written in the two-sublattice (Nambu) representation:
\be
H_0 = \sum_{\bf k} \Psi^{\dagger}_{\bf k} {\cal H}({\bf k}) \Psi_{\bf k},
\label{ham-nambu}
\ee
where 
\bea
\Psi_{\bf k} &=& \left(
\begin{array}{clcr}
\psi_a ({\bf k})\\
\psi_b ({\bf k})
\end{array}
\right), ~~~
{\cal H}({\bf k}) = \left(
\begin{array}{cc}
0 & t({\bf k})\\
t^*({\bf k}) & 0
\end{array}
\right), \nn\\
t({\bf k}) &=& 1 + 2\cos(k_x/2)\re^{\ri(\sqrt 3/2)k_y},
\eea
and the sum in (\ref{ham-nambu}) goes over 
the Brillouin zone. 
The spectrum has two degeneracy points (nodes) at ${\bf Q}_{1,2} = (\pm 4\pi/3,0)$.
Linearizing the noninteracting Hamiltonian near these points yields two cones
associated with (2+1)-dimensional massless Dirac fermions:
\bea
{\cal H}({\bf Q}_{1,2} + {\bf p}) = v(\tau_y p_y \mp \tau_x p_x), ~~~~v = \sqrt 3 t/2.
\eea

When a two-dimensional sheet of graphite is wrapped to produce an armchair nanotube,
$k_y$ gets quantized, and the lowest-energy subband correspoding to $k_y =0$ stays gapless.
The resulting problem is one-dimensional because the wave function does not depend on $y$.
Projecting the fermionic annihilation operators of the $a$ and $b$ sublattices 
onto the $k_y =0$ subspace, we get:
\bea
\left(
\begin{array}{c}
\psi_a ({\bf r})\\
\psi_b ({\bf r})
\end{array}
\right) \rightarrow \re^{\ri Q x}\left(
\begin{array}{c}
r_1 (x)\\
l_1 (x)
\end{array}
\right) + \re^{-\ri Q x}\left(
\begin{array}{c}
r_2 (x)\\
l_2 (x)
\end{array}
\right).
\eea

The effective 1D Hamiltonian is brought to its canonical diagonal form
\bea
H_0 = - \ri v \sum_{j=1,2} \int \rd x~ \left( R^{\dagger}_j \p_x R_j
-  L^{\dagger}_j \p_x L_j \right) 
\eea
by linear transformations:
\bea
&&r_1 = (R_1 + L_1)/\sqrt 2, ~~ l_1 = (- R_1 + L_1)/\sqrt 2\nonumber\\
&&r_2 = (R_2 - L_2)/\sqrt 2, ~~ l_2 = (R_2 + L_2)/\sqrt 2 \label{trans}
\eea
The structure of these transformations reflects the fact that
different Dirac points, $k_x =  Q$ and $k_x =  - Q$, are not associated
with fermions of different chiralities, as it is the case for standard chains and ladders;
instead each of these points is characterized by a pair of right $(R_j)$ and left $(L_j)$ fields. 
This is because the gapless spectrum of the armchair nanotube keeps the memory of the
two-cone Dirac structure of the dispersion law in the 2D graphite (in fact, 
the 1D spectrum is obtained from the 2D one by cutting the two
cones by the plane $k_y = 0$). For this reason, in case of nanotubes,
smooth components of the physical fields are contributed not only by
the diagonal ``currents'', $R^{\dagger}_j R_j$ and $L^{\dagger}_j L_j$, but also by
off-diagonal ``mass bilinears'', $R^{\dagger}_j L_j$ and $L^{\dagger}_j R_j$.

The chiral fermionic fields can be bosonized in terms of chiral bosonic fields $\Phi^{R,L}_{j\s}$:
\bea
&& \left(
\begin{array}{c}
R_{j\s} \\
L_{j\s}
\end{array}
\right) = \frac{\kappa_{j\s}}{\sqrt{2\pi \alpha}} e^{\pm i \sqrt{4\pi}\Phi^{R,L}_{j\s} },
~~~j=1,2, ~~\s = \pm 1, \label{bosonization}\\
&& [\Phi^R _{j\s}, \Phi^L _{j',\s'}] = \frac{\ri}{4} \delta_{jj'} \delta_{\s\s'}.
\nonumber
\eea
Here $\alpha$ is the short-distance cutoff in the bosonic theory, and 
$\kappa_{j\s}$ are Klein factors obeying the algebra
$
\{ \kappa_{j\s}, \kappa_{j'\s'} \} = \delta_{jj'} \delta_{\s\s'}
$. The product of the four Klein factors,
$\Gamma = \kappa_{1\up}\kappa_{1\down} \kappa_{2\up}\kappa_{2\down}$, satisfies
$\Gamma^2 =1$. Since $\Gamma$ is not a dynamical variable, we can conveniently choose
$\Gamma =1$.

In the bulk of this paper, we adopted the description in terms of four scalar fields,
$\Phi^{(\pm)}_c$, $\Phi^{(\pm)}_s$,
and their dual counterparts, $\Theta^{(\pm)}_c$, $\Theta^{(\pm)}_s$,
known from earlier studies of the two-channel Kondo problem \cite{kivelson}. These fields
describe the symmetric and antisymmetric charge excitations (equivalently, the ``charge'' and ``flavor'' modes),
\bea
\Phi^{(\pm)}_c = \frac{1}{2} \left( \Phi_{1\up} + \Phi_{1\down} \pm 
\Phi_{2\up} \pm \Phi_{2\down} \right),
\label{f1}
\eea
as well as the symmetric and antisymmetric spin excitations(or the ``spin'' and ``spin-flavor'' modes),
\bea
\Phi^{(\pm)}_s = \frac{1}{2} \left( \Phi_{1\up} - \Phi_{1\down} \pm 
\Phi_{2\up} \mp \Phi_{2\down} \right).
\label{f2}
\eea
Here $\Phi_{j\s} = \Phi^R _{j\s} + \Phi^L _{j\s}$. The corresponding dual fields,
$\Theta^{(\pm)}_c$ and $\Theta^{(\pm)}_s$ are obtained from the above expressions by
replacing $\Phi_{j\s}$ by $\Theta_{j\s} = - \Phi^R _{j\s} + \Phi^L _{j\s}$.

\section{}

The interaction can be written as 
\bea
&&[\rho_a + \rho_b]({\bf r}_1)V_{aa}({\bf r}_{12})[\rho_a + \rho_b]({\bf r}_2) + \nonumber\\
&&2\rho_a({\bf r}_1)[V_{ab}({\bf r}_{12}) - V_{aa}({\bf r}_{12})]\rho_b({\bf r}_2),\label{inter}
\eea
where, in the
low-energy limit, the local electron densities on the $a$ and $b$ sublattices are 
represented by 
\bea
&&\rho_a({\bf r}) \to  \rho_r (x) + \re^{2\ri Qx}M_r (x)  + \re^{-2\ri Qx}M^+_r (x), 
\nonumber\\
&&
\rho_b({\bf r}) \to \rho_l (x) + \re^{2\ri Qx}M_l (x) + \re^{-2\ri Qx}M^+_l (x),
\nonumber\\
&&
\rho_r = r_1^+r_1 + r_2^+r_2, ~~\rho_l = l^+_1l_1 + l^+_2l_2
\nonumber\\
&&
M_r = r^+_2r_1, ~~ M_l = l^+_2l_1 \label{dens}
\eea
Substituting Eq.(\ref{dens}) into Eq.(\ref{inter}) and dropping the oscillatory terms 
we get four terms:
\bea
&& \bullet ~~~ (\rho_r + \rho_l)_1V_{aa}({\bf r}_{12})(\rho_r + \rho_l)_2,\nn
&&\bullet ~~~ 2[U_{ab}(0) - U_{aa}(0)]\rho_r(x)\rho_l(x),\nn
&& \bullet ~~~ 2U_{aa}(2Q)[M_r(x)M_r^+(x) + M_l(x)M_l^+(x)],\nn
&& \bullet ~~~ 2U_{ab}(2Q)(M_rM_l^+ + M_lM_r^+),\nonumber
\eea
where $U(0)$ and $U(2Q)$ stand for the Fourier transforms of the interaction potentials.
 The first term here gives the Luttinger coefficient renormalization. 
The 'backscattering' interaction expressed in terms of the standard Dirac fermions with 
flavour indices 1,2 looks as follows:

\bea
&&\frac{1}{2}g_1(R_{1,\s}^+L_{1,\s} - R_{2,\s}^+L_{2,\s} + h.c.)^2 + \nonumber\\
&&g_2[(L^+_{2,\s}L_{1,\s})(R_{1,\s'}^+R_{2,\s'}) + h.c.] + \nonumber\\
&&g_3(L^+_{2,\s}R_{1,\s} - R^+_{2,\s}L_{1,\s})(L^+_{1,\s'}R_{2,\s'} -  R^+_{1,\s'}L_{2,\s'})
\label{interaction}
\eea
where the couplings are expressed in terms of the Fourier components of the interaction:
\bea
&&g_1 = U_{aa}(0) - U_{ab}(0), \nonumber\\
&&g_2 = U_{ab}(2Q) - U_{aa}(2Q), ~~ g_3 = U_{aa}(2Q) + U_{ab}(2Q).
\eea
In the UV model, which apart from the Hubbard (on-site) interaction $U$ also includes
the interaction between electrons on nearest-neighbor sites, $V$, the couplings $g_i$
are given by
\be
g_1 = U-3V, ~~~g_2 = -U, ~~~g_3 = U
\label{g-UV}
\ee

The bare Hamiltonian for $R,L$ is standard: 
\bea
H_0 = \ri v \int \rd x[L^+_a\p_x L_a - R^+_a\p_x R_a]
\eea

As already explained in section I and II, 
in the presence of long-range Coulomb potential,
Umklapp processes with the structure
$R^{\dagger}_1 R^{\dagger}_2 L_3 L_4 + h.c.$ represent the strongly relevant part
of interaction (\ref{interaction}). 
In the bosonic language, these are the processes containing the symmetric charge field
$\Phi^{+}_{\rc}$.
Using bosonization rules (\ref{bosonization}) -- 
(\ref{f2}), one straightforwardly derives Eq. (\ref{tunn}).
The $g_2$-part of (\ref{interaction}) does not contribute.

 If the interaction is the Coulomb one, then the  couplings $g_1,g_3$ 
are positive. It also looks likely that $g_3 > g_1$ and $|g_3-g_1| << g_{1,3}$.  
At these circumstances we have to consider two possibilities: (i) $g_1 < g_3$ 
(Haldane SL) and (ii) $g_1 > g_3$ (CDW)
.

\end{document}